\documentclass[traditabstract]{aa} % for the abstract without structuration 
\usepackage{graphicx}
\usepackage{txfonts}
\begin{document}
\title{Gamma-rays from the vicinity of accreting neutron stars inside compact 
high-mass X-ray binaries}

%   \subtitle{I. Overviewing the $\kappa$-mechanism}

   \author{W. Bednarek
%          \inst{1}
%          \and
%          C. Ptolemy\inst{2}\fnmsep\thanks{Just to show the usage
%          of the elements in the author field}
          }

 \institute{Department of Astrophysics, University of \L \'od\'z,
              90-236 \L \'od\'z, ul. Pomorska 149/153, Poland\\
             \email{bednar@fizwe4.phys.uni.lodz.pl}}
%   \institute{Institute for Astronomy (IfA), University of Vienna,
%              T\"urkenschanzstrasse 17, A-1180 Vienna\\
%              \email{wuchterl@amok.ast.univie.ac.at}
%         \and
%             University of Alexandria, Department of Geography, ...\\
%             \email{c.ptolemy@hipparch.uheaven.space}
%             \thanks{The university of heaven temporarily does not
%                     accept e-mails}

   \date{Received ; accepted }

% \abstract{}{}{}{}{} 
% 5 {} token are mandatory

\abstract{Dense wind of a massive star can be partially captured by a neutron star (NS) 
inside a compact binary system. Depending on the parameters of NS and the wind, the matter can penetrate the inner NS magnetosphere. At some distance from the NS a very turbulent and magnetized transition region is formed due to the balance between the magnetic pressure and the pressure inserted by accreting matter. This region provides good conditions for acceleration of particles to relativistic energies.
The matter at the transition region can farther accrete onto the NS surface (the accretor phase) or is expelled from the NS vicinity (the propeller phase). We consider the consequences of acceleration of electrons at the transition region concentrating on the situation in which at least part of the matter falls onto the NS surface. This matter creates a hot spot on the NS surface which emits thermal radiation.
Relativistic electrons lose energy on the synchrotron process and the inverse Compton (IC) scattering of this thermal radiation. We calculate the synchrotron spectra (from X-rays to soft $\gamma$-rays) and IC spectra (above a few tens MeV) expected in such a scenario. It is argued that a population of recently discovered massive binaries by the INTEGRAL observatory, which contain neutron stars hidden inside dense stellar winds of massive stars, can be detectable by the recently launched {\it Fermi} LAT telescope at  GeV energy
range. As an example, we predict the expected $\gamma$-ray flux from recently discovered source IGR J19140+0951.
}

\titlerunning{Gamma-rays from accreting neutron stars}

\maketitle

\section{Introduction}

Massive binary systems, containing a compact object (neutron star or black hole), have been suspected as sites of particle acceleration up to at least TeV energies since early 80'ties. But only recently, three such type of objects have been confirmed as VHE $\gamma$-ray sources by the Cherenkov telescopes. It is clear that at least in one case the $\gamma$-ray emission is due to acceleration of particles 
as a result of interaction of the energetic pulsar wind with the wind of a massive star.
Other TeV $\gamma$-ray binaries either operate under this same general scenario or the acceleration of particles occurs in the jet launched by the compact object. 
In the case of these TeV $\gamma$-ray binaries, the neutron star (pulsar) should rotate fast enough in order to produce energetic pulsar wind. 

In this paper we consider  the case when the neutron star
is not so energetic and the matter from the stellar wind can penetrate into the inner pulsar 
magnetosphere. At certain distance from the NS surface the magnetic pressure balances the 
pressure of accreting matter creating very turbulent transition region. Then, the NS is enshrouded in a dense cocoon and resembles objects called as the {\it hidden pulsars} (see e.g. Tavani~1991, Tavani \&  Brookshow~1993) or cauldrons (Begelman \& Rees~1984, Treves et al.~1993). Note that $\gamma$-ray emission has been predicted from such {\it hidden pulsars} (Tavani~1993). We consider the situation in which electrons are accelerated in turbulent, strongly magnetized, transition region in the inner pulsar magnetosphere. In principle, electrons can even reach $\sim TeV$ energies
but, as we show below, due to large synchrotron energy losses only $\gamma$-rays in the {\it Fermi} LAT telescope energy range may be effectively produced.
 
It looks that the INTEGRAL observatory has recently discovered a class of objects 
within the sample of compact high mass X-ray binaries (see e.g. Chaty et al~2007, Rodriquez \& Bodaghee~2008) which might operate under the mechanism discussed by us. These newly discovered massive binaries are compact, with the orbital periods a few to several days. Some of them contain relatively slowly rotating neutron stars which may allow the matter to penetrate the inner NS magnetosphere. According to the classification scheme of X-ray binaries (e.g. Lipunov~1992) such binaries with slowly rotating neutron stars (NS) belong to the propeller and accretor class. In these objects, the matter of the dense stellar wind accrete onto the strongly magnetized NS, interacting with the rigidly rotating inner
magnetosphere. In the case of accretor most of the matter reach the NS surface.
In the case of propeller most of the matter is expelled from the NS vicinity due to the centrifugal force.

It is proposed that such general scenario can be applied to a type of obscured compact objects inside massive binaries recently observed by the INTEGRAL observatory.  We perform detailed calculations of the $\gamma$-rays spectra including in some cases also the inverse Compton (IC) e$^\pm$ pair cascade initiated by relativistic electrons in the radiation field produced on the neutron star surface. We also take into account the synchrotron energy losses of primary electrons and cascade $e^\pm$ pairs, in order to obtain simultaneous synchrotron X-ray spectra. The acceleration of hadrons in such a scenario and possible production of high energy $\gamma$-rays, neutrinos and neutrons will be discussed in another paper.

\section{Description of the model}

\begin{figure}
\vskip 4.5truecm
\includegraphics{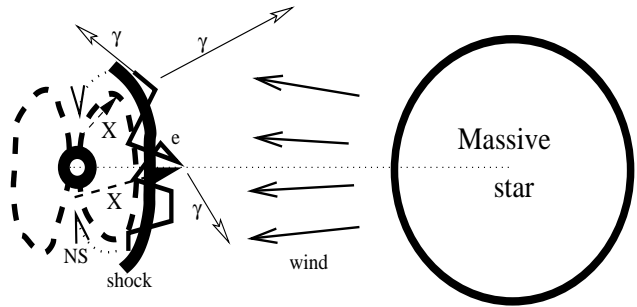}
\caption{Schematic representation of the model discussed in the paper.
The wind from a massive star penetrate the inner magnetosphere of a neutron star.
As a result of the interaction of the wind with the dipole magnetic field of the NS a strongly magnetized turbulent region is created (thick solid curve) in which the matter gains energy  being accelerated to the velocity of the rotating magnetic field lines. Electrons (e) are accelerated in this turbulent transition region and cool on the synchrotron process and on the comptonization of thermal radiation (marked by X), emitted from the stellar surface around the magnetic pole region. As a result, the first generation of IC $\gamma$-rays is produced. Thermal radiation appears due to the gravitational energy release of accreting matter onto the NS surface. If the optical depths for $\gamma$-rays are large enough, they initiate the IC e$^\pm$ pair cascade in this same thermal radiation.}
\label{fig1}
\end{figure}

We consider a compact binary system containing rotating neutron star 
(NS) and a massive companion of the O,B type star. It is assumed that a mass from the 
stellar wind is effectively captured by a strong gravitational potential of the NS.
Such slowly rotating neutron stars appear at a certain stage of the evolution of the binary system due to the loss of angular momentum either by the pulsar mechanism or 
as a result of the torque exhibited by the matter accreted from the stellar wind. 
Depending on the rotational period and surface magnetic field of NS, the accretion process onto NS can occur in different phases. According to the classification scheme of Lipunov~(1992), the accretion process can occur in the phase of accretor (for relatively slower rotators) or 
in the phase of propeller. For very energetic, strongly magnetized and short period neutron stars, the accretion process does not occur at all since the matter can not penetrate below the light cylinder radius. 
In the case of NS in the accretor and propeller phases, the matter from the stellar wind
penetrate below the light cylinder radius of the rotating NS magnetosphere. This matter 
extracts rotational energy from the NS as a result of the interaction of a free falling matter with the rigidly rotating inner NS magnetosphere. In this paper, we consider such direct collision of the  matter from the stellar wind with the NS magnetosphere (see general scenario shown in Fig.~\ref{fig1}). As a result of such interaction, a very turbulent and magnetized transition region is formed. In the case of accretor, most of the matter falls onto the NS surface creating small hot region on the NS surface.
In the case of propeller, most of the matter is expelled from the vicinity of the NS.
It is expected that the matter might be able to accrete also in this stage but in a rather non-stationary form since due to gradual accumulation of the matter close to the transition region the pressure of the matter can overcome the pressure from the rotating magnetosphere. It is supposed that some amount of the matter can accrete but most of it is expelled from the binary system (forming a larger scale jets ?) for some specific transition parameters describing the accretor/propeller scenario.

Note that in some massive binaries (e.g. Her X-1), the accretion of matter can occur also through the Roche lobe overflow. Then, the accretion disk around the NS is at first formed and the matter from the disk interacts directly with the NS magnetosphere. We do not discuss this possibility in this paper but rather concentrate on a simpler quasi-spherical accretion onto magnetized NS of the matter from the dense wind of the massive star.
The accretion through the disk can be easily taken into our considerations in our scenario by introducing a geometrical factor which describes the part of the sphere in which the accretion occurs. Therefore, the accretion through the disk will correspond to one of our discussed cases with effectively larger accretion rate occurring only inside a limited region determined by the thickness of the accretion disk.

Let us at first consider the case of the NS in the phase of pure accretor.
The accretion rate of the matter onto the neutron star (${\dot M}_{\rm acc} = 10^{16}M_{\rm 16}$ g s$^{-1}$) can be estimated from the observed (in some cases) thermal X-ray emission ($L_{\rm X} = 10^{36}L_{36}$ erg s$^{-1}$). These two values can by related for the known radius and the mass of NS (we assume $R_{\rm NS}\approx 10^6$ cm and $M_{\rm NS} = 1.4M_\odot$),
\begin{eqnarray}
{\dot M}_{\rm acc}\approx 5\times 10^{15}L_{\rm 36}~~~~{\rm g~s^{-1}}.
\label{eq1}
\end{eqnarray}
\noindent 
This matter arrives onto the neutron star surface along the magnetic field lines.
The distance at which the magnetic field starts to dominate the dynamics of the matter (the Alfven radius) can be estimated by comparing the magnetic field energy density with the kinetic energy density of the matter,

\begin{eqnarray}
B_{\rm A}^2/8\pi = \rho v_{\rm f}^2/2,
\label{eq2}
\end{eqnarray}
\noindent
where $B_{\rm A}$ is the magnetic field in the inner neutron star magnetosphere, 
$\rho = {\dot M}_{\rm acc}/(\pi R_{\rm A}^2v_{\rm f})$ is the density of accreting matter, $v_{\rm f} = (2GM_{\rm NS}/R_{\rm A})^{1/2}$ is the free fall velocity of accreting matter, $R_{\rm A}$ is the Alfven radius, and G is the gravitational constant. The medium in the transition region is very turbulent and strongly magnetized providing good conditions for the acceleration of particles to high energies. Let us estimate the location of this region from the surface of the neutron star by applying Eq. \ref{eq2} and assuming that the magnetic field in the neutron star magnetosphere is of the dipole type, i.e. $B_{\rm A} = B_{\rm NS} (R_{\rm NS}/R_{\rm A})^3$,
\begin{eqnarray}
R_{\rm A} = 4\times 10^8 B_{\rm 12}^{4/7}M_{\rm 16}^{-2/7}~~~~{\rm cm},
\label{eq3}
\end{eqnarray}
\noindent
where the magnetic field at the neutron star surface is $B_{\rm NS} = 10^{12}B_{12}$ (see also Baan \& Treves~1973). Based on the known vale of $R_{\rm A}$, we can estimate the magnetic field strength at the transition region,
\begin{eqnarray}
B_{\rm A} = 1.6\times 10^4M_{\rm 16}^{6/7}B_{\rm 12}^{-5/7}~~~~{\rm G}.
\label{eq4}
\end{eqnarray}
Considered here accretor scenario can occur provided that the neutron star fulfills some conditions. At first, the radius of the transition region has to 
lay inside the light cylinder radius of the neutron star, i.e $R_{\rm A} < R_{\rm LC} = cP/2\pi$, where $P = 1P_{1}$ s is the rotational period of the neutron star, and $c$ is the velocity of light. The above condition is fulfilled when, 
\begin{eqnarray}
P_{1} > 0.084  B_{\rm 12}^{4/7}M_{\rm 16}^{-2/7}.
\label{eq5}
\end{eqnarray}
\noindent
Therefore, only relatively slowly rotating neutron stars can be considered. 

At second, the rotational velocity of the magnetosphere at $R_{\rm A}$ has to be lower than the keplerian velocity of the accreting matter. The rotational velocity, 
\begin{eqnarray}
v_{\rm rot} = 2\pi R_{\rm A}/P\approx 2.5\times 10^9 B_{12}^{4/7} M_{16}^{-2/7}/P_{1}~~~{\rm  cm~s^{-1}}.
\label{eq6}
\end{eqnarray}
is lower than the keplerian velocity,
 \begin{eqnarray}
v_{\rm k} = (GM_{\rm NS}/R_{\rm A})^{1/2}\approx  6.8\times 10^8B_{12}^{-2/7} M_{16}^{1/7}~~~{\rm  cm~s^{-1}}.
\label{eq7}
\end{eqnarray}
for the NS with the periods,
\begin{eqnarray}
P_{1} > 3.7 B_{12}^{6/7}M_{16}^{-3/7}.
\label{eq8}
\end{eqnarray}
\noindent
This last condition separates the population of NS in the propeller phase (lower periods) from these ones in the accretor phase (larger periods). Note, that discussed below scenario for acceleration of electrons can be applied also to the propeller stage of accretion onto the NS. However, in the propeller phase the radiation field 
created by the matter accreting onto the NS is not uniquely defined. A part of the matter accreting onto NS can not be at present stage of knowledge linked to the basic parameters
describing the model. We have to introduce additional factor which describes the amount of the matter which falls onto the NS surface to the total amount of the matter penetrating 
the inner NS magnetosphere (i.e accreting and expelled from the vicinity of NS).

The third condition relates the radius of the transition region, $R_{\rm A}$, to the capturing radius of the matter from the stellar wind. It is determined by the balance between the kinetic energy of the wind with its gravitational energy around the NS. The capturing radius is described by, 
\begin{eqnarray}
R_{\rm c} = 2GM_{\rm NS}/v_{\rm w}^2\approx 3.7\times 10^{10}v_8^{-2}~~{\rm cm},  
\label{eq9}
\end{eqnarray}
\noindent
where $v_{\rm w} = 10^8v_8$ cm s$^{-1}$ is the velocity of surrounding matter measured in respect to the NS. The accretion from the stellar wind occurs when $R_{\rm c} >  R_{\rm A}$, which happens for, 
\begin{eqnarray}
B_{12} < 2.8\times 10^3v_8^{-7/2}M_{16}^{1/2}.  
\label{eq10}
\end{eqnarray}
For likely parameters of the neutron stars (classical and millisecond pulsars), this limit is not restrictive. Note that $v_{\rm w}$ corresponds to the velocity of the stellar wind and/or the velocity of the neutron star on its orbit around the massive star. This last velocity can be estimated from, $v_{\rm NS} = 2\pi D/T = 5\times 10^7 D_{2}/T_{10}$ cm s$^{-1}$, where $D = 100D_{2}R_\odot$ cm is the radius of the orbit in units of 100 solar radii $R_\odot$, and $T = 10T_{10}$ days is the period of the binary system in units of 10 days. It is typically lower than the free fall velocity at the $R_{\rm A}$. Therefore, we neglected it when deriving Eq.~\ref{eq3}.

Neutron stars with the periods within the range defined by Eq.~5 and Eq.~8 are in the propeller phase, those ones with the periods above that given by Eq.~8 are in the accretor phase, and those ones with the periods shorter than given by Eq.~5 are in the ejector phase.
As already noted, only in the accretor phase the accretion rate can be directly linked to the thermal X-ray emission from the NS surface. In the propeller phase, the amount of the matter which accrete onto the NS surface can not be uniquely determined from the basic 
parameters of the model.

\subsection{Acceleration of electrons}

In the conditions expected for the transition region (strongly magnetized and very turbulent medium), particles should be efficiently accelerated. In this paper we consider only acceleration of electrons. The energy gain rate of electrons with energy $E$ (and the Lorentz factor $\gamma$) is often parametrized by the Larmor radius of electrons and so called acceleration parameter,
\begin{eqnarray}
{\dot P}_{\rm acc} = \xi c E/r_{\rm L}
\approx 2.6\times 10^4\xi_{-1}M_{16}^{6/7}B_{12}^{-5/7} ~~~{\rm erg~s^{-1}},
\label{eq11}
\end{eqnarray}
\noindent
where $\xi = 10^{-1}\xi_{-1}$ is the acceleration parameter, $c$ is the velocity of light, 
$r_{\rm L} = E/eB_{\rm A}$ is the Larmor radius, and $e$ is the electron charge. 
The acceleration parameter contains all the unknown details of the acceleration process.
Since the plasma moves in the transition region with the velocity which takes a significant part of the velocity of light, it seems proper to consider the values of $\xi$ in the range $\sim 0.1-0.01$. During the acceleration process, electrons also suffer energy losses on the synchrotron process and the inverse Compton scattering of radiation from the massive star and the surface of the neutron star. These energy losses determine the maximum energies of accelerated electrons since their Larmor radius is typically much smaller than the characteristic dimensions of the considered scenario, i.e. $R_{\rm L} < R_{\rm A}$. This last condition allows in principle acceleration of electrons up to $E_{\rm e}\approx 2M_{16}^{4/7}B_{12}^{-8/7}$ PeV. However, as we show below, energy losses of electrons limit their maximum energies significantly below this value. 

Electrons lose energy on the IC process in the Thomson (T) and the Klein-Nishina (KN) regimes. In general, this process can occur on the stellar radiation and on the thermal X-ray radiation from the surface of the polar cap on the NS. Let us estimate the photon energy densities from the star ($\rho_\star$) and the polar cap ($\rho_{\rm cap}$) at the acceleration region,
\begin{eqnarray}
\rho_\star = {{4\sigma T_\star^4}\over{c}}\left({{R_\star}\over{D}}\right)^2
\approx 6.1\times 10^3T_4^4\left({{R_\star}\over{D}}\right)^2~~{\rm erg~cm^{-3}},
\label{eq12}
\end{eqnarray}
\noindent
(where $T_\star = 3\times 10^4 T_4$ K and $\sigma$ is the Stefan-Boltzmann constant) and, 
\begin{eqnarray}
\rho_{\rm cap} = {{4\sigma T_{\rm cap}^4}\over{c}} 
{{R_{\rm cap}^2}\over{R_{\rm A}^2}}
\approx 1.4\times 10^{8}M_{16}^{11/7}B_{12}^{-8/7}~~{\rm {erg~cm^{-3}}},
\label{eq13}
\end{eqnarray}
\noindent
where $T_{\rm cap} = 10^7T_7$ K is the temperature of the polar cap on the NS surface,  
$R_{\rm cap}$ is the polar cap radius on the NS surface on which the matter accrete.
In Eq.~\ref{eq13}, we have assumed that the observed X-ray thermal emission is re-radiated from the region of the polar cap according to $L_{\rm X} = \pi R_{\rm cap}^2\sigma T_{\rm cap}^4$. In fact, the emission from the polar cap region is well 
described by the Black Body spectrum (see e.g. Zane et al.~2000).
The radius of the polar cap region on the NS surface, on which the matter falls
and from which the thermal X-ray emission is emitted, can be estimated from (assuming dipole structure of the magnetic field),
\begin{eqnarray}
R_{\rm cap} = (R_{\rm NS}^3/R_{\rm A})^{1/2} \approx 5\times 10^{4}B_{12}^{-2/7}M_{16}^{1/7} ~~~{\rm cm}.
\label{eq20}
\end{eqnarray}
\noindent
Then, the surface temperature of the polar cap is,
\begin{eqnarray}
T_{\rm cap} = (L_{\rm X}/\pi R_{\rm cap}^2\sigma)^{1/4}\approx 4.7\times 10^7B_{12}^{1/7}M_{16}^{5/28}~~{\rm K}.
\label{eq21}
\end{eqnarray}
Note that, the energy density of radiation from the polar cap at the distance of the turbulent region ($R_{\rm A}$) can be explicitly expressed by the basic parameters of the model.

\begin{figure}
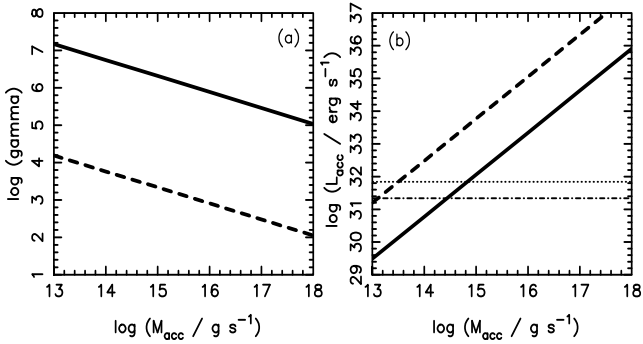

\vskip 4.5truecm
\includegraphics{nsfig2b.eps}
\includegraphics{nsfig2a.eps}
\caption{The maximum energies of accelerated electrons (a) and the maximum power which 
can be transfered to these electrons (b)  as a function of the accretion rate of the matter 
from the massive star onto the neutron star. The results are shown for the neutron stars which have the surface magnetic field $B = 10^{12}$ G (solid line), and $B = 10^9$ G (dashed). The acceleration parameter is equal to $\xi_{-1} = 1$. The estimated ${\it Fermi}$ LAT sensitivity at 1 GeV (1 yr) for the source at the distance of 1 kpc is $L_\gamma\approx 10^{32}$ erg s$^{-1}$ (thin dot-dashed line) and CTA sensitivity (50 hrs) at 20 GeV $L_\gamma\approx 3\times 10^{32}$ erg s$^{-1}$ (thin dotted line).}
\label{fig2}
\end{figure}

Let us also estimate the energy density of the magnetic field at the transition region ($R_{\rm A}$ given by Eq.~\ref{eq3}),
\begin{eqnarray}
\rho_{\rm B} = B^2_{\rm A}/8\pi
\approx 10^7 M_{16}^{12/7}B_{12}^{-10/7}~~{\rm {erg~cm^{-3}}},
\label{eq14}
\end{eqnarray}
\noindent
The energy losses of electrons for every considered process (synchrotron and IC in the T regime) can be calculated from,
\begin{eqnarray}
{\dot P}_{\rm loss} = (4/3)c\sigma_{\rm T}\rho\gamma^2\approx 2.7\times 10^{-14}\rho_{\rm (cap, B, \star)}\gamma^2~~{\rm erg~s^{-1}},
\label{eq15}
\end{eqnarray}
\noindent
where $\sigma_{\rm T}$ is the Thomson cross section. 

The energy losses of electrons on the radiation from the NS surface can dominate over  
energy losses on other targets only at low energies, i.e. when the scattering occurs in the T regime which happens for the Lorentz factors of electrons, 
$\gamma_{\rm T/KN}^{\rm X} < m_{\rm e}c^2/(3kT_\star)\approx 200T_7^{-1}$.
In the KN regime, the energy losses can be approximately estimated by introducing into Eq.~\ref{eq15} the value of the Lorentz factor corresponding to the transition between T and KN regimes, $\gamma_{\rm T/KN}^{\rm X}$. For the range of parameters defining the model, the IC energy losses of electrons in the KN regime becomes lower than the synchrotron energy losses for the Lorentz factors about ten times larger than $\gamma_{\rm T/KN}^{\rm X}$ (typically at $\sim$GeV energies). Therefore the maximum energies of accelerated electrons are determined by the balance between energy gains from the acceleration process (Eq.~\ref{eq11}) and energy losses on synchrotron process (Eq.~\ref{eq15}),
\begin{eqnarray}
\gamma_{\rm max}\approx 3\times 10^5 
\xi_{-1}^{1/2}B_{12}^{5/14}M_{16}^{-3/7}.
\label{eq16}
\end{eqnarray}
It is clear that electrons can reach even TeV energies (see Fig.~\ref{fig2}a) for realistic parameters of the model. However, TeV $\gamma$-ray production  will be strongly suppressed 
due to dominant synchrotron energy losses of electrons with TeV energies. 

We can also estimate the characteristic energies of synchrotron photons which can be expected in such a model by applying the approximate formula, 
\begin{eqnarray}
\epsilon_{\rm x}\approx m_{\rm e}c^2(B_{\rm A}/B_{\rm cr})\gamma_{\rm max}^2\approx  
1.9\xi_{-1}~~{\rm MeV},
\label{eq17}
\end{eqnarray}
\noindent
where $m_{\rm e}$ is the electron rest mass, $B_{\rm cr} = 4.4\times 10^{13}$ G is the critical magnetic field, and the Lorentz factor of electrons $\gamma_{\rm max}$ is given by Eq.~\ref{eq16}. We conclude that synchrotron emission can extend through the soft $\gamma$-ray energy range. it should be detectable by satellites sensitive at these energies, e.g. the INTEGRAL observatory.

Note that the bremsstrahlung energy losses of relativistic electrons in the matter inside the transition region can be neglected since their energy loss rate is relatively low,
\begin{eqnarray}
{\dot P}_{\rm br} = m_{\rm e}c^3\rho_{\rm H} \gamma/X_{\rm o}\approx 6.5\times 10^{-5}\rho_{17}\gamma~~{\rm erg~s^{-1}},
\label{eq18}
\end{eqnarray}
\noindent
where $X_{\rm o} = 62$ g cm$^{-2}$ is the radiation length for hydrogen, and $\rho_{\rm H} = 10^{17}\rho_{17}$ cm$^{-3}$ is the density of matter.
We can neglect the bremsstrahlung energy losses in respect to other energy loss rates since typical density of matter in the transition region is $\rho\approx 1.5\times 10^{13} B_{12}^{-6/7} M_{16}^{10/7}$ cm$^{-3}$ (estimated for the known accretion rate and the 
location of $R_{\rm A}$).

\subsection{Energetics}

The maximum power available for acceleration of electrons is limited by the energy 
extracted from the rotating neutron star by the in-falling matter. This matter from the stellar wind has to be accelerated to the velocity of the magnetic field lines at $R_{\rm A}$ in order to farther accrete onto the neutron star surface. The power which has to be transfered from the rotating NS to the matter can be estimated from
\begin{eqnarray}
L_{\rm acc} = {\dot M}_{\rm acc} v_{\rm rot}^2/2 \approx 3\times 10^{34}B_{12}^{8/7}M_{16}^{3/7}P_{1}^{-2}~~{\rm erg~s^{-1}}.
\label{eq19}
\end{eqnarray}
\noindent
By using the limiting period given by Eq.~\ref{eq8}, we get the upper limit on the available power in the accretor stage,
\begin{eqnarray}
L_{\rm acc} < 2.2\times 10^{33}B_{12}^{-4/7}M_{16}^{9/7}~~{\rm erg~s^{-1}}.
\label{eq20}
\end{eqnarray}
\noindent
This power is shown in Fig.~2b for two example surface magnetic field strengths of the neutron star (classical NS: $B = 3\times 10^{12}$ G and millisecond NS: $B = 10^9$ G) as a function of the accretion rate.
We assume that only a part, $\eta$, of this power (applying typically $\eta = 0.1$) can be converted to relativistic electrons in the turbulent and magnetized plasma at $R_{\rm A}$. 
Note, that for reasonable accretion rates this power is greater than the minimum required power of the GeV $\gamma$-ray source which can be detected by the ${\it Fermi}$ LAT telescope
and possibly also by the planned next generation Cherenkov telescopes system CTA whose sensitivity should be about one order of magnitude better than presently available.

The model discussed above for the accretor stage of the binary system can also operate for the propeller stage in which only a part of the matter can eventually accrete onto the 
surface of the NS. However, in this case it is difficult to estimate the power which can be 
transfered to relativistic particles due to the unknown observational signatures of the 
amount of matter which accumulates close to the shock region.

In summary, electrons can be accelerated even to TeV energies in the case of neutron stars inside X-ray binaries with relatively weak thermal X-ray emission 
from its surface in contrary to the powerful X-ray binaries in which only a few ten GeV electrons are expected. However, the situation is the opposite concerning the maximum power transfered to these relativistic electrons. Sources in which electrons are accelerated only to GeV energies should be powerful enough to be observable by the $\gamma$-ray telescopes.

\section{Production of radiation} 

We assume that electrons are accelerated in the turbulent transition region with the power law spectrum to maximum energies estimated from Eq.~\ref{eq16}. Eventually, the spectrum of electrons can be strongly peaked at the highest possible energies due to the synchrotron energy losses during acceleration process (the so called pile-up mechanism, see e.g. 
Protheroe~2004). As we have shown above, electrons lose energy on different radiation mechanisms but the dominant ones are:
the synchrotron process (dominates at the highest energies); and the ICS of thermal radiation from the NS polar cap (contributes to the high energies in the KN regime and dominates at lower energies in the T regime). 
We neglect the production of $\gamma$-rays by electrons in the acceleration region in the
scattering of stellar radiation since its energy density can be safely neglected in respect to the energy density of the magnetic field (at high energies) and the energy density of polar cap radiation (at low energies) (see Sect.~2.1).  
In order to check whether electrons can lose efficiently energy already in the transition region, we estimate their convection time scale with the matter falling onto the NS surface along the open magnetic field lines on,
\begin{eqnarray}
\tau_{\rm conv} = R_{\rm A}/v_{\rm rot}= P_1/2\pi ~~~{\rm s}.
\label{eq21}
\end{eqnarray}
\noindent
Let us compare this time scale with the radiation time scale, which (for electrons with low enough energies) is determined by the IC losses (in T regime) in the radiation field from the polar cap of the NS,
\begin{eqnarray}
\tau_{\rm rad} = m_{\rm e}\gamma/{\dot P}_{\rm rad}\approx 0.2 M_{16}^{-11/7}B_{12}^{8/7}\gamma^{-1}~~~{\rm s}
\label{eq22}
\end{eqnarray}
\noindent
From this comparison, we estimate lower limit on the Lorentz factor of electrons
which lose efficiently energy before being convected onto the NS surface on, 
\begin{eqnarray}
\gamma = 1.3 M_{16}^{-11/7}B_{12}^{8/7}P_1^{-1}.
\label{eq23}
\end{eqnarray}
\noindent
It is concluded that relativistic electrons lose their energy close to the acceleration place. 

In order to calculate the $\gamma$-ray spectra produced by electrons inside the transition region, we simulate the energy loss process of electrons in the radiation field of the polar cap and at the magnetic field of the transition region by applying the Monte Carlo method. It is assumed that the magnetic field is very turbulent in the transition region
(the distribution of electrons is isotropic) and that the radiation field comes from the region of the polar cap. For some range of model parameters, IC $\gamma$-rays can be farther absorbed in the radiation field of the polar cap. In this way the IC $e^\pm$ pair cascade
is initiating. In order to determine the conditions for the cascade process, in the next Section we calculate the optical depths for $\gamma$-ray photons in the radiation of the polar cap. We calculate also the optical depths for $\gamma$-rays in the radiation field of the massive companion in order to check whether this $\gamma$-ray absorption process should be also taken into account when evaluating the $\gamma$-ray spectra escaping from the binary system.

\begin{figure}
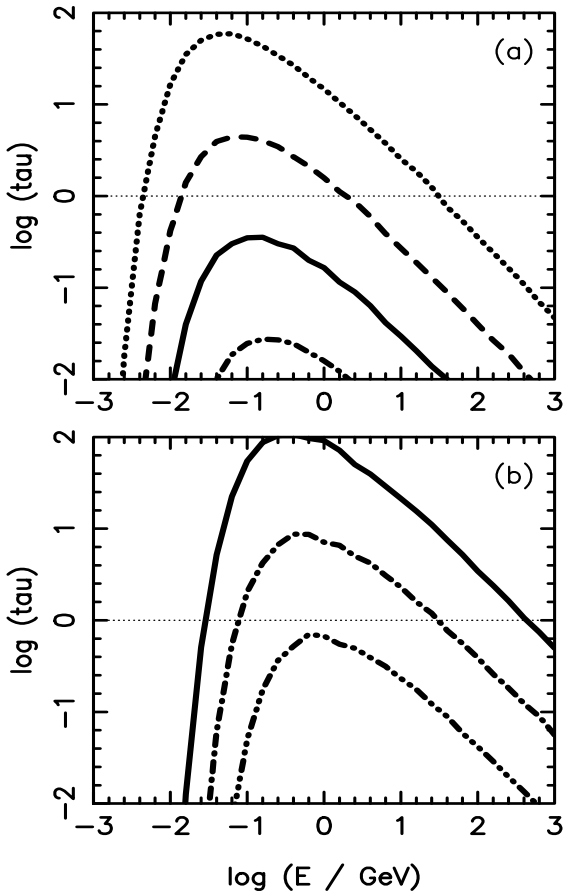

\vskip 12.truecm
\includegraphics{nsfig3a.eps}
\includegraphics{nsfig3b.eps}
\caption{The optical depths for $\gamma$-rays (averaged over the isotropic injection of $\gamma$-rays) in the thermal radiation from the polar cap on the NS surface as a function of $\gamma$-ray energies and for different accretion rates of matter onto the polar cap: ${\dot M}_{\rm acc} = 10^{18}$ g s$^{-1}$ (dotted curve), $10^{17}$ g s$^{-1}$ (dashed),  $10^{16}$ g s$^{-1}$ (solid), $10^{15}$ g s$^{-1}$ (dot-dashed),
and $10^{14}$ g s$^{-1}$ (triple dot-dashed).
The neutron star has the surface magnetic field equal to $B_{\rm NS} = 3\times 10^{12}$ G
(upper figure) and $10^9$ G (bottom). The distance of the acceleration region from the polar cap ($R_{\rm A}$) is defined by the above parameters. The thin dotted line marks the optical depth equal to unity.}
\label{fig3}
\end{figure}
\section{Optical depths for $\gamma$-rays}

As we have argued above, the cooling of electrons in such strong magnetic and radiation fields occurs very efficiently. High energy $\gamma$-ray produced in the ICS of thermal photons from the polar cap can be also absorbed in this same radiation field 
and, in principle, also in the radiation field of the close massive companion star. Below, we calculate the optical depths for the $\gamma$-ray photons in these two radiation fields. We show that for specific conditions (determined mainly by the accretion rate onto the NS and the parameters of the NS), the $\gamma$-ray spectrum emerging from the binary system is in fact formed in the cascade process occurring in the radiation field of the polar cap.

\subsection{Polar cap radiation}

We calculate the optical depths for $\gamma$-rays assuming that they are produced by electrons in the transition region at the distance $R_{\rm A}$ from the stellar surface. In these calculations we assume the diluted black body spectrum for the polar cap emission with the dilution factor
at the production site of $\gamma$-rays estimated from $(R_{\rm cap}/R_{\rm A})^2$. 
The average optical depths (averaged over the isotropic injection of $\gamma$-rays) for different accretion rates onto the neutron star (i.e equivalent to different X-ray luminosities, see Eq.~1) are shown in Fig.~\ref{fig3}.
It is clear that for low accretion rates ($<10^{16}$ g s$^{-1}$), when electrons can be accelerated to $\sim$TeV energies, the optical depths for $\gamma$-ray photons are relatively low, i.e. they are below unity for energies above $\sim 10$ GeV. In this case GeV-TeV $\gamma$-rays can escape from the radiation field of the polar cap without significant absorption.

On the other hand, for large accretion rates ($>10^{16}$ g s$^{-1}$), electrons are accelerated at most to a few tens of GeV. $\gamma$-rays produced by these electrons should be  efficiently absorbed in the radiation of the polar cap. Therefore, in this case the escaping $\gamma$-ray spectra can be only obtained by calculating complicated IC $e^\pm$ pair cascade in the radiation of the polar cap. 
We conclude that for large accretion rates $\gamma$-rays, produced in this model, could be observable by the satellite telescopes (${\it Fermi}$ LAT and possibly AGILE).

\subsection{Massive star radiation}

\begin{figure}
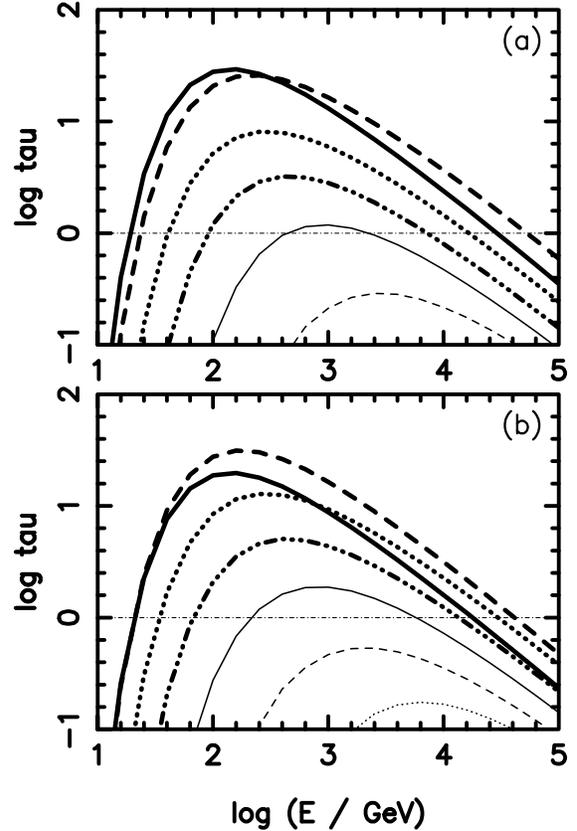

\vskip 11.5truecm
\includegraphics{nsfig4b.eps}
\includegraphics{nsfig4a.eps}
\caption{The optical depths as a function of $\gamma$-ray energies are shown for the case of the massive star in the massive binary IGR J19140+098 discovered by INTEGRAL telescope: $R_\star = 30R_\odot$, $T_\star = 2.8\times 10^4$ K and two distances from the star $D = 3R_\star$ (a) and $2R_\star$ (b). Specific curves show the optical depths for the injection angles $\alpha = 0^{\rm o}$
(thick solid curve), $30^{\rm o}$ (thick dashed), $60^{\rm o}$ (thick dotted), $90^{\rm o}$ (thick dot-dot-dashed), $120^{\rm o}$ (thin solid), $150^{\rm o}$ (thin dashed), and $180^{\rm o}$ (thin dotted), measured from the direction towards the star.}
\label{fig4}
\end{figure}

The optical depths for $\gamma$-ray photons injected at an arbitrary distance from the surface of the massive star have been calculated for the first time in the general case (including also dimensions of the star) by Bednarek (1997, 2000). In fact, the optical depths for $\gamma$-rays close to the surface of other stars can be easily re-scaled from those early calculations. For example, $\gamma$-ray photons
with energies, $E_\gamma^{\rm o}$, propagating at specific
distances $D$ and at directions (defined by the angle $\alpha$) close to the star 
with specific parameters ($T_{\rm o}$ and $R_{\rm o}$) are related to the optical depths 
around arbitrary star with $T_\star$ and $R_\star$ in the following way,
\begin{eqnarray}
\tau(E_\gamma^\star={{E_\gamma^{\rm o}}\over{S_{\rm T}}},T_\star,R_\star,D,\alpha) = 
S_{\rm T}^3S_{\rm R}\tau(E_\gamma^{\rm o}, T_{\rm o}, R_{\rm o},D,\alpha)
\label{eq24}
\end{eqnarray}
\noindent
where $S_{\rm T} = T_\star/T_{\rm o}$, $S_{\rm R} = R_\star/R_{\rm o}$, distance $D$ from the stellar surface of specific star is measured in the stellar radii. The example calculations of the optical depths for the case of one specific massive star (present inside the binary system IGR J19140+098) are shown in Fig~\ref{fig4}. These optical depths become larger than unity for the $\gamma$-rays with energies above $\sim$20 GeV. As we show latter, $\gamma$-rays produced in the considered model have typically energy below $\sim$10 GeV. Their absorption in the radiation field of the massive star can be safely neglected. Therefore, the absorption effects of $\gamma$-rays should not introduce any modulation of the $\gamma$-ray signal with the period of the binary in the case of accreting neutron stars. This is in contrast to the TeV $\gamma$-ray binaries supposed to contain ejecting neutron stars (strong pulsar wind prevents the accretion of stellar wind), see e.g. the model considered by Sierpowska \& Bednarek~(2005). The eventual modulation of the $\gamma$-ray signal with the period of the binary can be only produced by the change of the accretion rate with the distance of the neutron star on its elliptic orbit around the massive star.

\section{Gamma-rays from the vicinity of accreting neutron star} 

In the considered model, radiation is produced by electrons
accelerated to maximum energies estimated by Eq.~\ref{eq16} in the synchrotron and IC process. The synchrotron energy losses dominate at the largest electron energies (close to $E_{\rm max}$) and the IC process can only dominate at lower energies due to the KN cross section. However, as we have shown above, for large accretion rates, $\gamma$-rays produced by electrons can be farther absorbed in the thermal radiation from the NS surface. 
We have shown in Fig.~\ref{fig3} that the average optical depths for $\gamma$-ray photons in the thermal radiation from the NS polar cap region in order to envisage for what parameters the cascading effects can become important.
Therefore, in these cases we have to consider the production of the $\gamma$-rays in the IC $e^\pm$ pair cascade with additional synchrotron energy losses of primary electrons and secondary $e^\pm$ pairs. On the other hand, for small accretion rates, the cascade does not develop. $\gamma$-rays are produced in this case only as a result of cooling of primary electrons. In order to follow the process of $\gamma$-ray production, we developed the numerical code which simulate the 
cooling process of electrons taking into account not only the $\gamma$-ray production in the IC process but also the synchrotron energy losses of primary electrons and secondary cascade $e^\pm$ pairs. Note that the acceleration of electrons occurs inside the inner pulsar magnetosphere where the magnetic field is relatively strong and depends on the distance of the turbulent transition region from the NS surface. Since the synchrotron process dominates over the IC process at the high energy part of the injected electron spectrum, 
the $\gamma$-rays are rarely produced with energies comparable to energies of primary electrons. Our IC $e^\pm$ pair cascade code with synchrotron energy losses gives both, the synchrotron and the IC spectra expected in such a model.

Two models for injection of relativistic electrons are considered:

\begin{enumerate}

\item Electrons injected with the power law spectrum up to the maximum 
Lorentz factor $\gamma_{\rm max}$ (see Eq.~\ref{eq16}), as expected in the 
stochastic acceleration mechanism. The power law spectrum of primary electrons 
is normalized to a part, $\eta$, of the kinetic power which is transfered from the rotating neutron star to the matter in the transition region.

\item Electrons injected with the Lorentz factors $\gamma_{\rm max}$. Such injection spectrum can give good approximation in the case of acceleration process occurring with large radiative energy losses (the pile up mechanism at the end of accelerated power law spectrum of electrons, see e.g. Protheroe~2004).

\end{enumerate}

These two injection models are likely to be the limiting cases of the real acceleration
process of electrons in the turbulent region of the matter accreting onto the magnetized neutron star.

We investigate two general scenarios describing the cases of {\it pure} accretor phase
(all the matter arriving to the transition region falls onto the NS surface) and the intermediate accretor-propeller phase (only part of the matter arriving to the transition region accrete onto the NS surface and the rest of it is expelled from the vicinity of the NS in the propeller mechanism).

\subsection{The accretor stage}

\begin{figure*}[h]
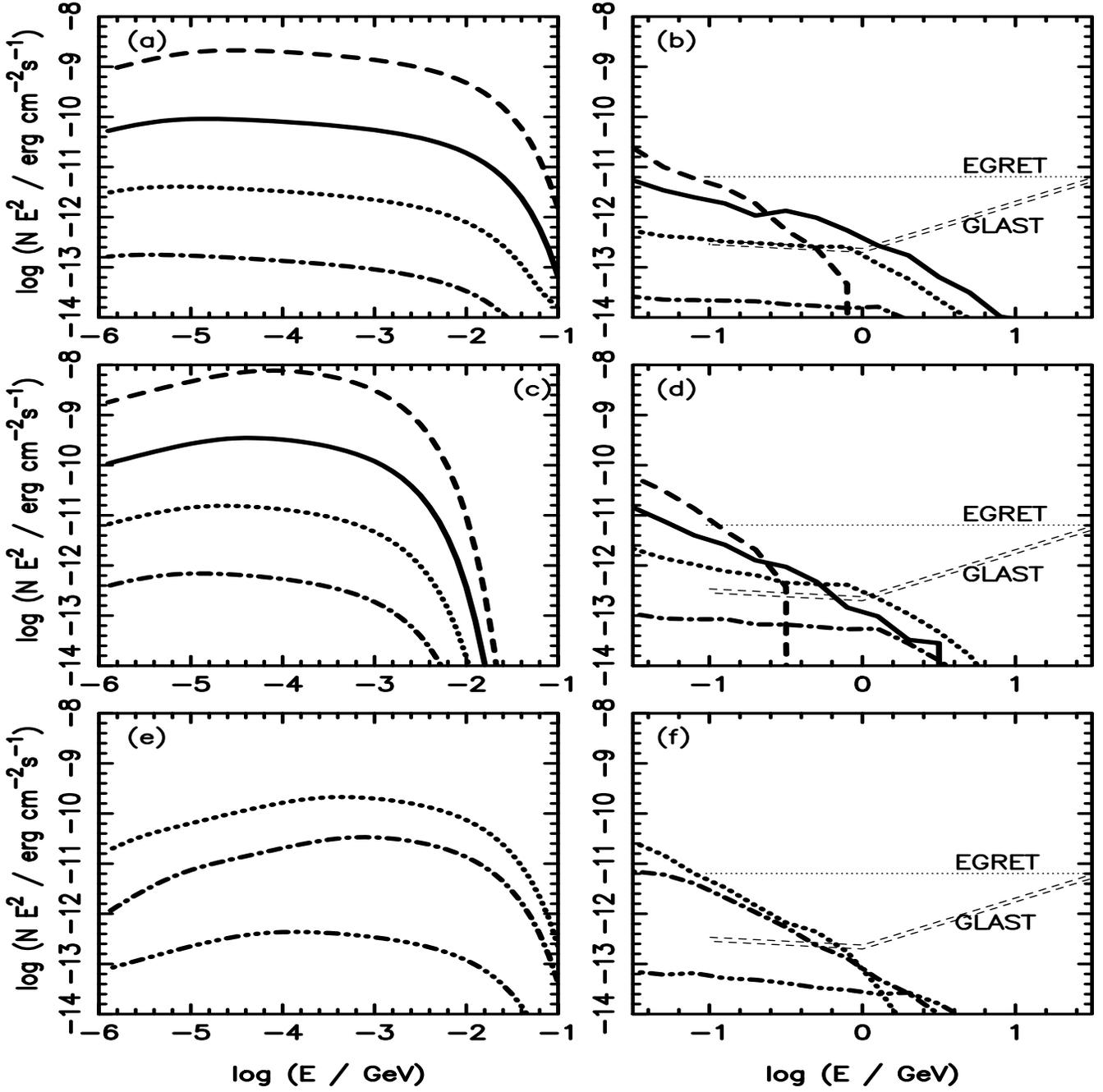

\vskip 18.truecm
\includegraphics{nsfig5a.eps}
\includegraphics{nsfig5b.eps}
\includegraphics{nsfig5c.eps}
\includegraphics{nsfig5d.eps}
\includegraphics{nsfig5e.eps}
\includegraphics{nsfig5f.eps}
\caption{The IC $\gamma$-ray spectra (on the right) and synchrotron spectra (on the left)
produced in a sequence of radiation processes by relativistic electrons. The IC process 
occurs on  the thermal radiation field produced by the matter accreting onto the neutron 
star surface. The X-ray and  $\gamma$-ray spectra are shown for different 
accretion rates: $M = 10^{18}$ g s$^{-1}$ (dashed curve), $10^{17}$ g s$^{-1}$ (solid), 
$10^{16}$ g s$^{-1}$ (dotted), $M_{\rm acc} = 10^{15}$ g s$^{-1}$
(dot-dashed), $10^{14}$ g s$^{-1}$ (triple dot-dashed). The spectra are shown for the acceleration parameter $\xi = 0.1$, the surface magnetic field of the NS $B_{\rm NS} = 10^{12}$ G (upper figures),  $\xi = 0.01$ and $B_{\rm NS} = 10^{11}$ G (middle), and  $\xi = 0.1$ and $B_{\rm NS} = 10^{9}$ G (bottom).  The spectrum of accelerated electrons is of the power law type ($dN(E)/dE\propto E^{-s}$) with the spectral index $s = 2.2$ between 30 MeV and the maximum energy given by Eq.~\ref{eq16}. The factor determining the power transfered from the transition region to relativistic electrons depends on the energy conversion efficiency, $\eta$, and the distance to the  source, $D = 1D_1$ kpc. It is assumed to be equal to $\eta D_1^{-2} = 0.1$. The period of the pulsar is equal to the limiting value given by Eq.~\ref{eq8}. The level of sensitivity of the the EGRET and GLAST telescopes are marked by thin dotted and double-dashed lines.}
\label{fig5}
\end{figure*}

The scenario for the accretion process onto neutron star in the accretor stage is much better defined since the thermal radiation field from the NS surface (depending on the dimension of the hot spot and its temperature) is uniquely determined by the parameters  describing the model.
Therefore, the synchrotron and IC spectra produced by electrons depends on a relatively small number of free parameters, i.e. the accretion rate onto NS, the surface magnetic field of NS, and the acceleration rate $\xi$. The observed X-ray to $\gamma$-ray power
depends additionally on the parameter, $\eta D^{-2}$, which combines the energy conversion efficiency from the transition region to relativistic electrons, $\eta$, and the distance to the source $D = 1D_1$ kpc. 
At first, we performed calculations for the power law spectrum of injected electrons 
$N(e) = AE^{-s}$ between $E_{\rm max}$ and $E_{\rm min}$, where $A$ is the
normalization coefficient equal to 
$A = \eta L_{\rm acc}(2-s)/(E_{\rm max}^{(2-s)} - E_{\rm min}^{(2-s)})$.
Note, that $A$ only weakly depends on $E_{\rm min}$ for the spectral indices not far 
from 2. In our example calculations, we apply the spectral index equal to 2.2 and choose 
$E_{\rm min} = 30$ MeV since we cool electrons only to such minimum energies. Electrons with such energies are not able to produce $\gamma$-rays detectable by the Fermi LAT.   
The calculated spectra are also shown for different values of ${\dot M}$, $\xi$ and $B$ (see Fig.~\ref{fig5}).

In Figs.~\ref{fig5}a-d, we consider the case of classical relatively young  NS ($B\gg 10^9$ G), rotating with the limiting period defined by Eq.~\ref{eq8}. When confronting with the sensitivity of the {\it Fermi} LAT telescope, it becomes clear that $\gamma$-ray emission from such objects can be detected at energies below $\sim 1$ GeV in the case of large accretion rates (significantly above $10^{16}$ g s$^{-1}$). Moreover, the power law synchrotron spectra, corresponding to these accretion rates, should be also easily detected by the soft and hard X-ray detectors presently on the orbit. Note that due to strong magnetic field and large energies of primary electrons, the synchrotron spectra can extend in some cases also through the soft $\gamma$-ray energy range. 
In Fig.~\ref{fig5}e-f, we show the synchrotron and IC spectra for the case of the NS with low surface magnetic field, i.e. characteristic for so called millisecond pulsars. Also in the case of these slowly magnetized neutron stars, $\gamma$-ray emission can be detected by the {\it Fermi} LAT telescope, provided that the accretion rate onto the NS is above a few $10^{14}$ g s$^{-1}$. We conclude that $\gamma$-ray telescopes presently on the orbit should be able to detect $\gamma$-ray emission at a few hundred MeV from the class of accreting NS. These NS are characterized by the 
thermal emission from its surface at a level above $\sim 2\times 10^{36}$ erg s$^{-1}$ in the case of strongly magnetized neutron stars, and above a few $10^{34}$ erg s$^{-1}$ in the case of NS with the magnetic field characteristic for millisecond pulsars provided that the source is at the distance of 1 kpc. These limiting X-ray luminosities have been derived by using the relation between the accretion rate and the thermal emission from the NS surface 
(see Eq.~\ref{eq1}).

\begin{figure*}[h]
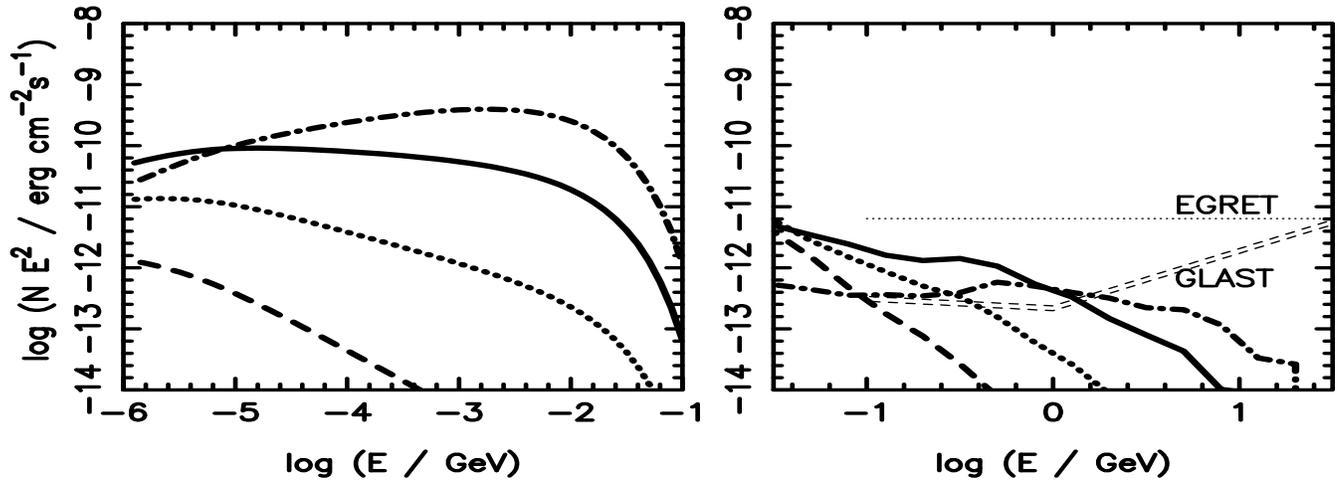

\vskip 6.5truecm
\includegraphics{nsfig6a.eps}
\includegraphics{nsfig6b.eps}
\caption{As in Fig.~\ref{fig5} but for different spectral indexes of injected electrons:
$s = 4$ (dashed), 3 (dotted), 2.2 (solid), 1 (dot-dashed).  The parameters describing the model are the following: $\xi = 0.1$, the surface magnetic field of the NS $B_{\rm NS} = 10^{12}$ G, and the accretion rate $M_{\rm acc} = 10^{17}$ g s$^{-1}$, .}
\label{fig6}
\end{figure*}

We also investigate how the detectability of such accreting NSs depends on the spectral index of the injected electrons. In Fig.~\ref{fig6} the synchrotron and IC $\gamma$-ray spectra are shown for the range of spectral indexes $s = 1-4$ and for fixed other parameters describing the model: ${\dot M} = 10^{17}$ g s$^{-1}$,
$\xi = 0.1$, and $B = 10^{12}$ G. Electron spectrum has to have spectral index lower than 
$\sim 3$ in order to produce IC $\gamma$-rays above the sensitivity of ${\it Fermi}$ LAT telescope. Such spectral indexes are expected in the stochastic models
for particle acceleration in turbulent media. 

\begin{figure*}[h]
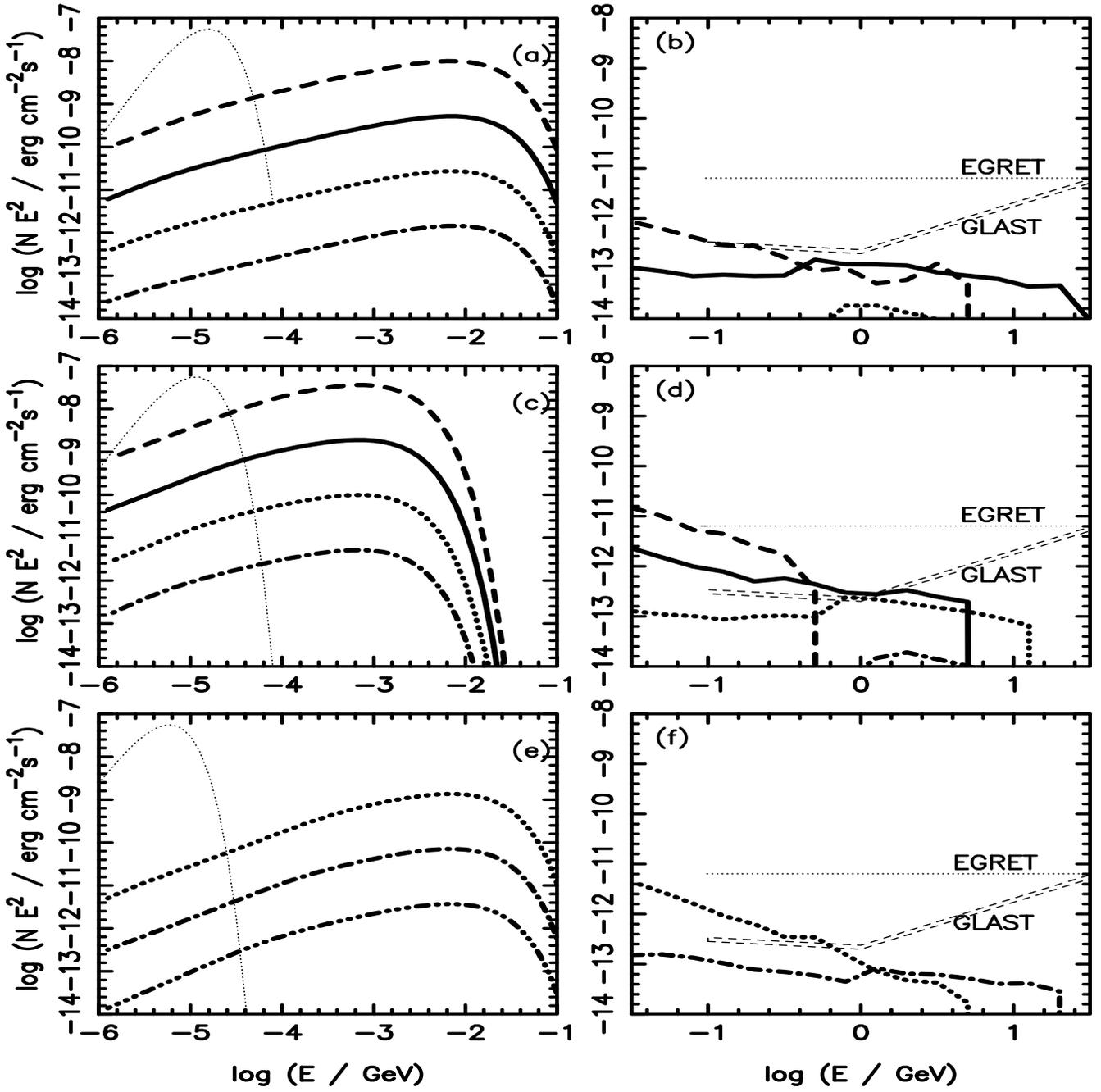

\vskip 18.truecm
\includegraphics{nsfig7a.eps}
\includegraphics{nsfig7b.eps}
\includegraphics{nsfig7c.eps}
\includegraphics{nsfig7d.eps}
\includegraphics{nsfig7e.eps}
\includegraphics{nsfig7f.eps}
\caption{As in Fig.~\ref{fig5} but for mono-energetic injection of electrons with energies
 given by Eq.~\ref{eq16}. The specific curves show the results obtained for these same parameters as in Fig.~\ref{fig5}. The spectrum of thermal radiation emitted from the 
neutron star surface is also shown for the typical accretion rate 
$M_{\rm acc} = 10^{16}$ g s$^{-1}$ and corresponding surface magnetic field strengths (thin dotted curves in (a), (c), and (e). }
\label{fig7}
\end{figure*}

As already noted above, the acceleration of particles under strong radiative energy losses may result in their accumulation at the highest energies determined by the balance between energy gains from the acceleration mechanism and energy losses (see e.g. Protheroe~2004). 
In such cases, most of the energy is accumulated in particles with the highest possible energies. At first approximation, we can consider that the spectrum of particles is mono-energetic.
Therefore, in Fig.~\ref{fig7} we show also the photon spectra expected in the model for the case of mono-energetic injection of electrons. In such a case, IC $\gamma$-ray spectra are flat. Their detectability is more difficult than in the case of injection of electrons with the power law spectra. This is due to the fact that electrons with maximum possible energies lose most of their energy on synchrotron process. Therefore the IC spectra are reduced and synchrotron spectra are enhanced (compare corresponding cases shown in Figs.~\ref{fig5} and~\ref{fig7}).
We conclude that accreting NS, which have the highest chance to be detected by the {\it Fermi} LAT telescope, should have the intermediate values of $\xi$ and $B$ and large accretion rates (see Fig.~\ref{fig7}c-d). Such parameters allow acceleration of electrons to energies of the order of a few tens GeV, i.e close to the energy region where ICS
process starts to become comparably efficient to the synchrotron process.

The $\gamma$-ray photons produced by electrons in the transition region of the accretion flow onto the NS originate close to the massive companion star which creates strong radiation field. In Sect. 4.2, we show the optical depths for $\gamma$-rays in the radiation field of the massive star with the parameters typical for massive binary systems (the example case of the massive binary IGR J19140+0951).
As we have shown above,  primary electrons accelerated to the maximum energies cool at first mainly on the synchrotron process. Therefore $\gamma$-rays produced by them in the IC process have energies typically below $\sim 10$ GeV.
On the other hand, the optical depths in the massive star radiation (see Fig.~\ref{fig4}) are quite large but at energies which are clearly above energies of $\gamma$-rays produced in our model. Therefore, we conclude that in the case of considered here model the absorption of $\gamma$-rays in the stellar radiation can be neglected.

In Fig.~7a,c,e, we also show the spectra of thermal radiation from the NS polar cap 
region for the example accretion rate $M_{\rm acc} = 10^{16}$ g s$^{-1}$ and different values of the magnetic field on the NS surface ($B = 10^{12}$ G (a), $10^{11}$ G (c),
and $10^{9}$ G (e). These spectra clearly dominate over the nonthermal synchrotron spectra
produced by relativistic electrons below a few keV. In fact, as we already noted above,
such soft X-ray thermal excesses have been reported from some INTEGRAL hard X-ray massive binaries. In reality this soft thermal X-ray emission may be significantly modified due to the interaction with the matter accreting onto the NS surface, the matter of the stellar
wind, and the matter onto the surface of the close massive companion. In our calculations 
we are not able to take these possible modifications into account.

\subsection{The intermediate accretor-propeller stage}

As noted above, in the {\it pure} propeller stage the matter arriving to the transition region at the Alfven radius, $R_{\rm A}$, is expelled by the centrifugal source from the vicinity of the neutron star preferentially along the rotational axis.
Here we consider the intermediate case in which a small part of the accreting matter is able to penetrate onto the NS surface but most of it is expelled outside the neutron star.
It is not clear whether such process can last stationary in time or it occurs only non-stationary when the accretion rate onto the neutron star changes in time e.g. due to the clumpy wind or the elliptic orbit of the NS around the massive star. 
In order to take into account the effects of only partial accretion of matter onto the NS surface, we introduce the parameter, $\kappa$, which is the ratio of the matter accreting onto the NS surface to the whole amount of matter arriving to the transition region (i.e. 
the matter accreted onto the surface and expelled from the vicinity of NS).
In such an intermediate case, the temperature of the polar cap region on the NS surface (given by Eq.~\ref{eq21}) should be a factor $\kappa^{1/4}$ lower than estimated in the case of complete accretion. However, in principle the power available for acceleration of electrons can be larger than expected from period limiting the accretor and propeller stage (given by Eq.~\ref{eq8}). Instead, it is rather limited by Eq.~\ref{eq5}. 
In conclusion, as a result of only partial accretion of matter from the transition region
onto the NS surface, the surface temperature of the polar cap is lower and the power transfered to electrons can be larger than expected for the {\it pure} accretor stage.

\begin{figure*}[t]
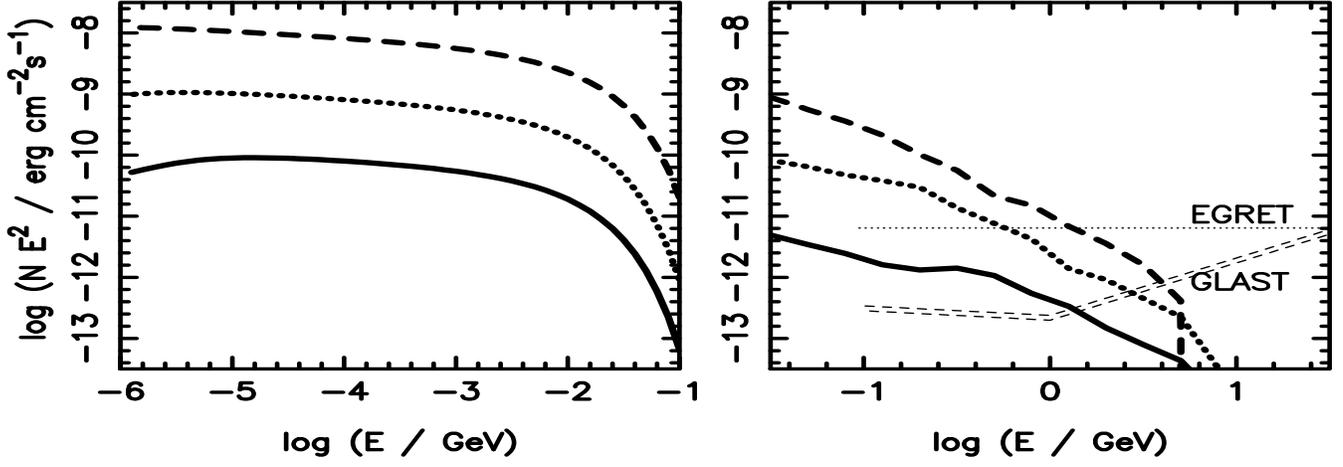

\vskip 6.5truecm
\includegraphics{nsfig8a.eps}
\includegraphics{nsfig8b.eps}
\caption{As in Fig.~\ref{fig5} but for different rates describing the amount of the matter accreting onto the NS surface in respect to the whole amount of the matter 
arriving to the transition region (accreting and expelled from the vicinity of the NS in propeller mechanism): $\kappa = 1$ (solid curve), $0.1$ (dashed), and $0.01$ (dotted). The basic parameters of the model are the following: ${\dot M} = 10^{17}$ g s$^{-1}$, $B = 10^{12}$ G, and $\xi = 0.1$.}
\label{fig8}
\end{figure*}

In Fig.~\ref{fig8} we show the example $\gamma$-ray spectra obtained in the case of the 
partial accretion of matter onto the NS surface assuming different accretion rates defined by the factor $\kappa$. We also scale the power transfered to relativistic electrons by the factor $\kappa^{-1}$ since more energy can be extracted from the transition region in this case. Based on these calculations, we expect that the $\gamma$-ray spectra 
produced in the intermediate stage are steeper, although they have larger luminosities. Therefore,  they should be easier detected by the $\gamma$-ray telescopes (see Fig.~\ref{fig8} for the comparison with the sensitivity of ${\it Fermi}$ LAT telescope for the source at the distance of 1 kpc).

\section{Discussion and Conclusions}

We propose that accreting neutron stars inside compact massive binary systems produce
$\gamma$-ray fluxes which can be observable by the satellite telescope ${\it Fermi}$ LAT.
To show this we consider acceleration of electrons at the turbulent, strongly magnetized, transition region in the inner neutron star magnetosphere which appears as a result of the interaction of accreting matter with rotating NS magnetosphere. Relativistic electrons 
produce X-rays and $\gamma$-rays as a result of synchrotron and IC processes occurring on the thermal radiation from the NS surface. 
The cooling process of electrons is followed by applying the Monte Carlo method since some of produced $\gamma$-rays can be farther absorbed in the thermal radiation.
We showed that synchrotron emission from primary electrons and secondary cascade $e^\pm$ pairs can extend up to the MeV energy range and the IC emission can appear up to a few GeV. 

\begin{figure*}
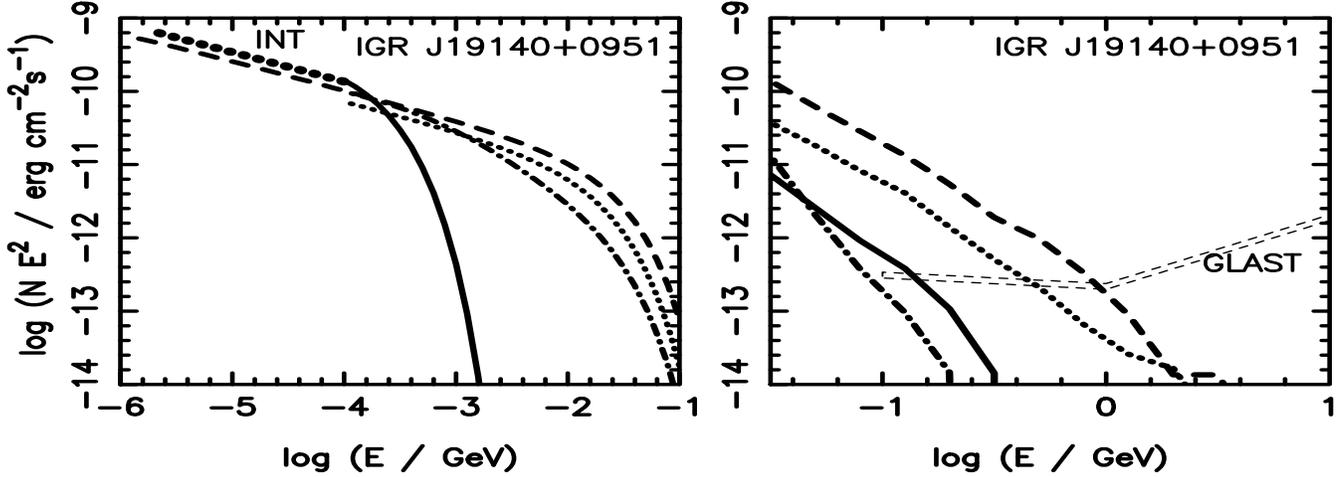

\vskip 6.5truecm
\includegraphics{nsfig9a.eps}
\includegraphics{nsfig9b.eps}
\caption{(a) The comparison of the hard X-ray spectra detected by the INTEGRAL observatory from the massive binary IGR J19140+0951 with the synchrotron spectra expected from the vicinity of neutron star for the pure accretor model with the parameters: ${\dot M} = 2\times 10^{17}$ g s$^{-1}$, $B = 10^{11}$ G, and $\xi = 0.1$, spectral index of electrons $s = 2.8$ (solid curve), ${\dot M} = 3\times 10^{17}$ g s$^{-1}$, $B = 10^9$ G, and $\xi = 0.1$, $s = 4$ (dot-dashed),  the intermediate accretor-propeller models for: ${\dot M} = 10^{17}$ g s$^{-1}$, $B = 10^{11}$ G, $\xi = 0.1$, and $\kappa = 0.1$ and $s = 2.8$ (dotted),  and ${\dot M} = 10^{17}$ g s$^{-1}$, $B = 10^{12}$ G, $\xi = 0.1$, and $\kappa = 0.01$ and $s = 2.8$ (dashed).  
(b) The IC $\gamma$-ray spectra expected in the model for the parameters mentioned above based on the normalization of the synchrotron spectra to the INTEGRAL observations in the X-ray energy range. The level of sensitivity of the the {\it Fermi} LAT telescope is marked by a double-dashed line.}
\label{fig9}
\end{figure*}

A physical realization of such a scenario can be characteristic for the newly discovered by the INTEGRAL observatory a class of obscured massive binary systems which show: (1) a hard, power law X-ray emission and, (2) evidences of soft X-ray black body component. 
As an example, we consider here in a more detail the high mass X-ray binary system, IGR J19140+0951, discovered by the INTEGRAL (Hannikainen et al.~2003). A compact object in this system is likely to be a neutron star (Cabanac et al.~2005).
IGR J19140+0951 is at the distance of $\sim 2-3$ kpc (Rahoui et al.~2008). The hard X-ray emission up to 100 keV with the spectral index $2.39\pm 0.11$ is observed from this source (Hannikainen et al.~2004). Its X-ray luminosity in the high state was $\sim 3.7\times 10^{37}\times (D/10 kpc)^2$ erg s$^{-1}$ (Rodriquez et al.~2005).
Unfortunately, the basic parameters of the neutron star in IGR J19140+0951 are unknown.
Therefore, we investigate the range of model parameters which are consistent with the hard 
power law spectrum in the X-ray energy range by the INTEGRAL. 
Comparison of the observations with the example calculations are shown in Fig.~\ref{fig9}.
It is clear that for some parameters the $\gamma$-ray flux predicted at a few hundred MeV can be detectable by the extensive {\it Fermi} LAT observations. 

We performed calculations for the neutron stars with the parameters characteristic for the 
"classical" radio pulsars (surface magnetic field strength $B\sim 10^{12}$ G and periods of 
the order of seconds) and for the millisecond pulsars ($B\sim 10^9$ G, periods of a few to several milliseconds). It is clear that in order to produce $\gamma$-ray fluxes observable by the ${\it Fermi}$ LAT telescope, the neutron stars in binary systems at the distance of a few kpc should collect the matter from the wind at a relatively large accretion rates 
($>10^{16}$ g s$^{-1}$). However, in the case of millisecond pulsars these accretion rates can be significantly lower (above a few $10^{14}$ g s$^{-1}$). Therefore, in principle,
the millisecond pulsars seems to be more favorite $\gamma$-ray sources. On the other hand, millisecond pulsars have only low mass companions, so the large accretion rates are not expected in such binary systems.

Note, that synchrotron X-ray and IC $\gamma$-ray emission is produced in such a model 
almost isotropically. As we have shown above, we do not expect any modulation of the $\gamma$-ray signal with the period of the binary system due to the selective absorption in the companion star soft radiation, as it is expected in the case of TeV $\gamma$-ray binary systems (e.g. LS 5039 and LSI 61 303). However, the modulation might be related to the change of the accretion rate in the case of elliptic orbit of the neutron star.
It is expected that the accretion rate should increase when the neutron star is closer to 
the companion star. Therefore, we predict that the largest fluxes of X-rays and $\gamma$-rays should be observed closer to the periastron passage of the neutron star.
Moreover, the X-ray and $\gamma$-ray emission should be correlated
(see spectra in Figs.~5-8). 

In the model considered here we do not take into account the scattering of nonthermal
synchrotron radiation by the accelerated electrons. As we have shown in Fig.~7, the thermal
radiation clearly dominates at energies below a few keV over the nonthermal synchrotron radiation produced by accelerated electrons and secondary cascade $e^\pm$ pairs. Since their energy densities at the acceleration site scales in a similar way with the distance from
the NS, the energy losses on thermal radiation has to dominate over the energy losses on nonthermal radiation.
Therefore, we can safely neglect the cooling of electrons on this nonthermal radiation. Note moreover, that the scattering of nonthermal X-rays with energies above a few keV becomes inefficient since it occurs in the Klein-Nishina regime.

Similar processes to these considered in this paper are also expected in the case of non-spherical accretion of matter onto the rotating neutron star. For example, in the case of disk accretion, the corresponding accretion rate should be scaled by a part of sphere which is intercepted by the accretion disk. Then, the turbulent region on the inner disk age, in which acceleration of electrons occurs, is also limited to corresponding part of the sphere.
Only the IC process may become more complicated in this case since thermal radiation from the hot spot on the neutron star surface, visible by relativistic electrons, can be partially obscured by the neutron star surface.

In conclusion, we predict that present satellite telescopes can discover a new class of $\gamma$-ray sources, i.e. accreting neutron stars inside binary system. The emission from these sources should be characterized by a strong hard, synchrotron spectra and IC $\gamma$-rays limited to energies below $\sim 1$ GeV, due to the very strong synchrotron energy losses of relativistic electrons accelerated in the vicinity of neutron stars.

\begin{acknowledgements}
This work is supported by the Polish MNiSzW grant N N203 390834. 
\end{acknowledgements}

%%%%%%%%%%%%%%%%%%%%%%%%%%%%%%%%%%%%%%%%%%%%

\end{document}